\documentclass[11pt]{article}
\usepackage{amssymb,amsmath,amsfonts}
\usepackage{graphicx}
\usepackage{graphics}
\usepackage{eepic,epsfig}

\begin{document}

\title{Casimir effect for parallel metallic plates\\
in cosmic string spacetime }
\author{E. R. Bezerra de Mello$^{1}$\thanks{%
E-mail: emello@fisica.ufpb.br}, \thinspace\ A. A. Saharian$^{1,2}$\thanks{%
E-mail: saharian@ysu.am}, \thinspace\ A. Kh. Grigoryan$^{3}$ \\
\\
\textit{$^{1}$Departamento de F\'{\i}sica, Universidade Federal da Para\'{\i}%
ba}\\
\textit{58.059-970, Caixa Postal 5.008, Jo\~{a}o Pessoa, PB, Brazil}\vspace{%
0.3cm}\\
\textit{$^2$Department of Physics, Yerevan State University,}\\
\textit{1 Alex Manoogian Street, 0025 Yerevan, Armenia} \vspace{0.3cm}\\
\textit{$^3$School of Physics and Mathematics, Yerevan State University,}\\
\textit{Azatutyan Ave., 0037 Yerevan, Armenia}} \maketitle

\begin{abstract}
We evaluate the renormalized vacuum expectation values (VEVs) of electric
and magnetic fields squared and the energy-momentum tensor for the
electromagnetic field in the geometry of two parallel conducting plates on
background of cosmic string spacetime. On the base of these results, the
Casimir-Polder force on a polarizable particle and the Casimir forces acting
on the plates are investigated. The VEVs are decomposed into the pure string
and plate-induced parts. The VEV of the electric field squared is negative
for points with radial distance to the string smaller than the distance to
the plates and positive for the opposite situation. On the other hand the
VEV for the magnetic field squared is negative everywhere. The boundary
induced part in the VEV of the energy-momentum tensor is different from zero
in the region between the plates only. Moreover, this part only depends on
the distance from the string. The boundary-induced part in the vacuum energy
density is positive for points with distance to the string smaller than the
distance to the plates and negative in opposite situation. The Casimir
stresses on the plates depend non-monotonically on the distance from the
string. We show that the Casimir forces acting on the plates are always
attractive.
\end{abstract}

\bigskip

PACS numbers: 03.70.+k, 98.80.Cq, 11.27.+d

\bigskip

\section{Introduction}

The Casimir effect is a phenomenon common to all systems characterized by
fluctuating quantities on which external boundary conditions are imposed
(for a review see \cite{Casimir}). It is among the most striking macroscopic
manifestations of the nontrivial properties of the quantum vacuum. The
boundary conditions imposed on a quantum field lead to the modification of
the spectrum for zero-point fluctuations. As a result of this, the
expectation values for physical quantities bilinear in the field are
shifted. In particular, the confinement of quantum fluctuations causes
forces that act on constraining boundaries. These forces depend on the
nature of the quantum field, on the bulk and boundary geometries, and on the
specific boundary conditions on the field operator.

Among the most interesting topics in the investigations of the Casimir
effect is the dependence of physical characteristics on the geometry of the
background spacetime. Closed analytic expressions can be obtained for highly
symmetric geometries only. In particular, motivated by Randall-Sundrum-type
braneworld scenarios, investigations of the Casimir effect in anti-de Sitter
spacetime have attracted a great deal of attention. The Casimir energy and
corresponding Casimir forces for two parallel branes in anti-de Sitter
spacetime have been evaluated in Refs. \cite{Noji00} by using either
dimensional or zeta function regularization methods. Local Casimir densities
were considered in Refs. \cite{Knap04}. The Casimir effect in
higher-dimensional generalizations of the anti-de Sitter spacetime with
compact internal spaces has been investigated in \cite{Flac03}. Another
popular background in gravitational physics is de Sitter spacetime. The
Casimir densities for a massive scalar field with general curvature coupling
parameter induced by flat and spherical boundaries on this background have
been recently investigated in \cite{Saha09} and \cite{Milt11}, respectively
(see also \cite{Seta01} for the case of a conformally coupled massless
scalar field). In the present paper we consider the Casimir effect for the
electromagnetic field in the geometry of two parallel conducting plates in
background of cosmic string spacetime.

Cosmic strings are topologically stable defects which may have been created
by phase transitions in the early Universe \cite{Vile94}. They are
candidates for the generation of a variety of interesting physical effects,
including gravitational lensing, anisotropies in the cosmic microwave
background radiation, the generation of gravitational waves, high-energy
cosmic rays, and gamma ray bursts. Recently the cosmic strings attracted a
renewed interest related to the appearance of a variant of their formation
mechanism within the framework of brane inflation \cite{Sara02}. In the
simplest theoretical model describing the infinite straight cosmic string
the spacetime is locally flat except on the string. In quantum field theory
the corresponding non-trivial topology induces non-zero vacuum expectation
values (VEVs) for physical observables. In this context, the VEVs of the
energy--momentum tensor have been evaluated for scalar, fermionic, and
electromagnetic fields \cite{Hell86}-\cite{Site11}. The analysis of the
Casimir effect in the cosmic string spacetime have been developed for scalar
\cite{Beze06}, fermionic \cite{Beze08,Beze10}, and electromagnetic fields
\cite{Brev95,Beze07} for the geometry of a coaxial cylindrical boundary. The
Casimir force for massless scalar fields subject to Dirichlet and Neumann
boundary conditions in the setting of the conical piston has been recently
investigated in \cite{Fucc11}. The Casimir densities for a scalar field
induced by a flat boundary perpendicular to the string have been considered
in \cite{Beze11}.

In the present paper we evaluate the VEVs of the electric and magnetic field
squared and the energy-momentum tensor for two parallel conducting plates
perpendicular to the cosmic string axis. These quantities are among the most
important local characteristics of the electromagnetic vacuum. Though the
corresponding operators are local, due to the global nature of the vacuum
state, they contain an important information about the topology of the
background spacetieme. In addition, the VEV of the energy-momentum tensor
plays an important role in modelling a self-consistent dynamics involving
the gravitational field. It is worth calling attention to the fact that the
renormalized VEV of the energy-momentum tensor for the electromagnetic field
in the geometry of a cosmic string without boundaries is evaluated in \cite%
{Frol87,Dowk87}. In the problem under consideration all calculations can be
performed in a closed form and it constitutes an example in which the
topological and boundary-induced polarizations of the vacuum can be
separated in different contributions.

The paper is organized \ as follows. In the next section we consider the
mode functions for the electric and magnetic fields in the region between
two conducting plates. In section \ref{sec:F2}, these mode functions are
used for the evaluation of the VEVs of the electric and magnetic fields
squared. By making use of the Abel-Plana summation formula, the latter are
decomposed as the sum of boundary-free, single plate-induced, and second
plate-induced parts. The Casimir-Polder force on a polarizable particle is
discussed as well. The VEV of the energy-momentum tensor and the Casimir
forces acting on the plates are investigated in section \ref{sec:EMT}.
Finally, in section \ref{sec:Conc} the main results are summarized.

\section{Mode functions for the electromagnetic field}

\label{sec:Inter}

In this investigation we consider the effect of two parallel metallic plates
on the quantum fluctuations of electromagnetic field in background of cosmic
string spacetime. For an infinitely long straight cosmic string the
corresponding line element in cylindrical coordinates with the string along
the $z$-axis is given by the expression%
\begin{equation}
ds^{2}=dt^{2}-dr^{2}-r^{2}d\phi ^{2}-dz{}^{2},  \label{ds21}
\end{equation}%
where $0\leqslant \phi \leqslant \phi _{0}$ and the spatial points $(r,\phi
,z)$ and $(r,\phi +\phi _{0},z)$ are to be identified. The planar angle
deficit is related to the mass per unit length of the string, $\mu _{0}$, by
$2\pi -\phi _{0}=8\pi G\mu _{0}$, where $G$ is the Newton gravitational
constant. We assume that the plates are orthogonal to the string and are
located at $z=0$ and $z=a$. In this paper we are interested in the change of
the vacuum expectation values (VEVs) of the electric and magnetic fields
squared and the energy-momentum tensor of the electromagnetic field, induced
by the plates.

The VEV for a physical quantity $f\left\{ F_{i},F_{k}\right\} $, bilinear in
the electric ($F=E$) or magnetic ($F=B$) fields, can be written as the
mode-sum%
\begin{equation}
\langle 0|f\left\{ F_{i},F_{k}\right\} |0\rangle =\sum_{\alpha }f\left\{
F_{\alpha i},F_{\alpha k}^{\ast }\right\} ,  \label{VEVbil}
\end{equation}%
where $\left\{ F_{\alpha i},F_{\alpha k}^{\ast }\right\} $ represents a
complete set of normalized mode functions, specified by a set of quantum
numbers $\alpha $, and obeying the boundary conditions%
\begin{equation}
\mathbf{n}\times \mathbf{E}_{\alpha }=0,\quad \mathbf{n}\cdot \mathbf{B}%
_{\alpha }=0,\quad z=0,a,  \label{BC}
\end{equation}%
where $\mathbf{n}$ is the normal vector to the plates (directed along the $z$
axis). We will consider the VEVs in the region between the plates, $%
0\leqslant z\leqslant a$. The VEVs for the regions $z\leqslant 0$ and $%
z\geqslant a$ are obtained as limiting cases.

We have two classes of mode functions corresponding to the waves of the
transverse magnetic (TM) and transverse electric (TE) types. In the case of
TM waves the corresponding mode functions for the electric field, obeying
the boundary conditions (\ref{BC}) on the plate at $z=0$, are given by the
expressions%
\begin{eqnarray}
E_{\alpha 1}^{(0)} &=&-\beta _{\alpha }k\gamma J_{q|m|}^{\prime }(\gamma
r)\sin (kz)e^{i\left( qm\phi -\omega t\right) },  \notag \\
E_{\alpha 2}^{(0)} &=&-i\beta _{\alpha }k\frac{qm}{r}J_{q|m|}(\gamma r)\sin
(kz)e^{i\left( qm\phi -\omega t\right) },  \notag \\
E_{\alpha 3}^{(0)} &=&\beta _{\alpha }\gamma ^{2}J_{q|m|}(\gamma r)\cos
(kz)e^{i\left( qm\phi -\omega t\right) },  \label{ETM}
\end{eqnarray}%
where $J_{\nu }(x)$ is the Bessel function, $0\leqslant \gamma <\infty $, and%
\begin{equation}
q=2\pi /\phi _{0},\;\omega ^{2}=\gamma ^{2}+k^{2},\;m=0,\pm 1,\pm 2,\ldots .
\label{omega}
\end{equation}%
In (\ref{ETM}), $E_{\alpha l}^{(0)}$ is the $l$-th physical component of the
electric field vector in cylindrical coordinates and the values $l=1,2,3$
correspond to the $r,\phi ,z$ coordinates, respectively. For the mode
functions corresponding to the magnetic field we find
\begin{eqnarray}
B_{\alpha 1}^{(0)} &=&\beta _{\alpha }\omega \frac{qm}{r}J_{q|m|}(\gamma
r)\cos (kz)e^{i\left( qm\phi -\omega t\right) },  \notag \\
B_{\alpha 2}^{(0)} &=&i\beta _{\alpha }\omega \gamma J_{q|m|}^{\prime
}(\gamma r)\cos (kz)e^{i\left( qm\phi -\omega t\right) },  \label{BTM}
\end{eqnarray}%
and $B_{\alpha 3}^{(0)}=0$. The eigenvalues for $k$ are quantized by the
boundary conditions (\ref{BC}) on the plate at $z=a$:%
\begin{equation}
k=k_{n}=\frac{\pi n}{a},\quad n=0,1,2,\ldots .  \label{kn}
\end{equation}

In the case of the TE waves, the mode functions have the form%
\begin{eqnarray}
E_{\alpha 1}^{(1)} &=&-\beta _{\alpha }\omega \frac{qm}{r}J_{q|m|}(\gamma
r)\sin (kz)e^{i\left( qm\phi -\omega t\right) },  \notag \\
E_{\alpha 2}^{(1)} &=&-i\beta _{\alpha }\omega \gamma J_{q|m|}^{\prime
}(\gamma r)\sin (kz)e^{i\left( qm\phi -\omega t\right) },  \notag \\
E_{\alpha 3}^{(1)} &=&0,  \label{ETE}
\end{eqnarray}%
for the electric field and%
\begin{eqnarray}
B_{\alpha 1}^{(1)} &=&\beta _{\alpha }k\gamma J_{q|m|}^{\prime }(\gamma
r)\cos (kz)e^{i\left( qm\phi -\omega t\right) },  \notag \\
B_{\alpha 2}^{(1)} &=&i\beta _{\alpha }k\frac{qm}{r}J_{q|m|}(\gamma r)\cos
(kz)e^{i\left( qm\phi -\omega t\right) },  \notag \\
B_{\alpha 3}^{(1)} &=&\beta _{\alpha }\gamma ^{2}J_{q|m|}(\gamma r)\sin
(kz)e^{i\left( qm\phi -\omega t\right) }.  \label{BTE}
\end{eqnarray}%
for the magnetic field, with the same notations as in (\ref{ETM}). Now for
the eigenvalues of $k$ we have $k=k_{n}=\pi n/a$, with $n=1,2,\ldots $. As
it is seen from the formulas for the mode functions, they are specified by
the set $\alpha =(\lambda ,\gamma ,m,n)$, where $\lambda =0,1$ corresponds
to the TM and TE waves, respectively.

The mode functions are normalized by the condition
\begin{equation}
\int_{0}^{\infty }drr\int_{0}^{\phi _{0}}d\phi \int_{0}^{a}dz\,\mathbf{E}%
_{\alpha }^{(\lambda )}\cdot \mathbf{E}_{\alpha ^{\prime }}^{(\lambda )\ast
}=2\pi \omega \delta _{\alpha \alpha ^{\prime }},  \label{NC}
\end{equation}%
where $\delta _{\alpha \alpha ^{\prime }}$ is understood as the Dirac delta
function for continuous components of the collective index $\alpha $ and as
the Kronecker delta for discrete ones. Substituting the expressions for the
mode functions for the electric field, it can be seen that the normalization
coefficient is given by the expression
\begin{equation}
\beta _{\alpha }^{2}=\frac{2q(1-\delta _{n0}/2)}{\omega \gamma a},
\label{betalf}
\end{equation}%
for the both TM and TE modes.

\section{VEVs of the electric and magnetic field squared}

\label{sec:F2}

In this section we consider the VEVs of the electric and magnetic fields
squared for the physical situation specified in the previous section. First
we consider the region between the plates, $0\leqslant z\leqslant a$.
Substituting the mode functions into the corresponding mode-sum formula (\ref%
{VEVbil}), for these VEVs we find%
\begin{eqnarray}
\langle 0|F^{2}|0\rangle &=&\sum_{\alpha }\mathbf{F}_{\alpha }^{(\lambda
)}\cdot \mathbf{F}_{\alpha }^{(\lambda )\ast }=\frac{4q}{a}%
\sideset{}{'}{\sum}_{m=0}^{\infty }\int_{0}^{\infty }d\gamma \frac{\gamma }{%
\omega }\left[ G_{qm}(\gamma r)\right.  \notag \\
&&\left. \times \sum_{n=1}^{\infty }\left( 2k_{n}^{2}+\gamma ^{2}\right)
f_{1}^{(F)}(k_{n}z)+\gamma ^{2}J_{qm}^{2}(\gamma r)\sideset{}{'}{\sum}%
_{n=0}^{\infty }f_{2}^{(F)}(k_{n}z)\right] ,  \label{F2}
\end{eqnarray}%
where $F=E$ and $F=B$ for the electric and magnetic fields respectively. In (%
\ref{F2}) we have defined the functions%
\begin{eqnarray}
f_{1}^{(E)}(x) &=&f_{2}^{(B)}(x)=\sin ^{2}x,  \notag \\
f_{2}^{(E)}(x) &=&f_{1}^{(B)}(x)=\cos ^{2}x,  \label{f2E}
\end{eqnarray}%
and%
\begin{equation}
G_{qm}(x)=J_{qm}^{\prime 2}(x)+\left( qm/x\right) ^{2}J_{qm}^{2}(x).
\label{Gqm}
\end{equation}%
Of course, the expressions (\ref{F2}) are divergent. We assume that a cutoff
function is introduced to make them convergent without explicitly writing
it. The special form of this function will not be important in the following
discussion.

For the evaluation of the sum over $n$ we apply the Abel-Plana summation
formula, written in the form (see, for example, \cite{SahaBook}):%
\begin{equation}
\frac{\pi }{a}\sideset{}{'}{\sum}_{n=0}^{\infty }f(\pi n/a)=\int_{0}^{\infty
}dx\,f(x)+i\int_{0}^{\infty }dx\,\frac{f(ix)-f(-ix)}{e^{2ax}-1},  \label{APF}
\end{equation}%
where the prime on the summation sign means that the term with $n=0$ should
be taken with the coefficient 1/2. This leads to the following
representation of the VEVs:%
\begin{equation}
\langle 0|F^{2}|0\rangle =\langle F^{2}\rangle _{1}+\langle F^{2}\rangle
_{2},  \label{F21}
\end{equation}%
where the first and second terms in the right-hand side correspond to the
first and second integrals in (\ref{APF}), respectively. For these separate
terms we get the expressions%
\begin{eqnarray}
\langle F^{2}\rangle _{1} &=&\frac{4q}{\pi }\sideset{}{'}{\sum}%
_{m=0}^{\infty }\int_{0}^{\infty }dk\int_{0}^{\infty }d\gamma \frac{\gamma }{%
\omega }[G_{qm}(\gamma r)  \notag \\
&&\times \left( 2k^{2}+\gamma ^{2}\right) f_{1}^{(F)}(kz)+\gamma
^{2}J_{qm}^{2}(\gamma r)f_{2}^{(F)}(kz)],  \label{F211}
\end{eqnarray}%
and
\begin{eqnarray}
\langle F^{2}\rangle _{2} &=&\frac{8q}{\pi }\sideset{}{'}{\sum}%
_{m=0}^{\infty }\int_{0}^{\infty }d\gamma \gamma \int_{\gamma }^{\infty }dx%
\frac{\left( x^{2}-\gamma ^{2}\right) ^{-1/2}}{e^{2ax}-1}  \notag \\
&&\times \left[ G_{qm}(\gamma r)\left( 2x^{2}-\gamma ^{2}\right)
g_{1}^{(F)}(xz)+\gamma ^{2}J_{qm}^{2}(\gamma r)g_{2}^{(F)}(xz)\right] ,
\label{F22}
\end{eqnarray}%
with the functions%
\begin{eqnarray}
g_{1}^{(E)}(x) &=&-g_{2}^{(B)}(x)=\sinh ^{2}x,  \notag \\
g_{2}^{(E)}(x) &=&-g_{1}^{(B)}(x)=\cosh ^{2}x.  \label{g2E}
\end{eqnarray}%
The term $\langle F^{2}\rangle _{1}$ does not depend on the distance between
the plates, whereas the term $\langle F^{2}\rangle _{2}$ vanishes in the
limit $a\rightarrow \infty $. From here it follows that the part $\langle
F^{2}\rangle _{1}$ corresponds to the VEV in the geometry of a single plate
located at $z=0$ when the second plate is absent. The part $\langle
F^{2}\rangle _{2}$ is induced by the second plate at $z=a$.

The single plate parts can be further decomposed as%
\begin{equation}
\langle F^{2}\rangle _{1}=\langle F^{2}\rangle ^{(s)}+\langle F^{2}\rangle
_{1}^{(b)},  \label{F1dec}
\end{equation}%
where%
\begin{eqnarray}
\langle E^{2}\rangle ^{(s)} &=&\langle B^{2}\rangle ^{(s)}=\frac{2q}{\pi }%
\sideset{}{'}{\sum}_{m=0}^{\infty }\int_{0}^{\infty }dk\int_{0}^{\infty
}d\gamma   \notag \\
&&\times \frac{\gamma }{\omega }\left[ G_{qm}(\gamma r)\left( 2k^{2}+\gamma
^{2}\right) +\gamma ^{2}J_{qm}^{2}(\gamma r)\right] ,  \label{E2s}
\end{eqnarray}%
is the corresponding VEV for a boundary-free string geometry. The terms%
\begin{eqnarray}
\langle E^{2}\rangle _{1}^{(b)} &=&-\langle B^{2}\rangle _{1}^{(b)}=-\frac{2q%
}{\pi }\sideset{}{'}{\sum}_{m=0}^{\infty }\int_{0}^{\infty }dk\cos (2kz)
\notag \\
&&\times \int_{0}^{\infty }d\gamma \frac{\gamma }{\omega }\left[
G_{qm}(\gamma r)\left( 2k^{2}+\gamma ^{2}\right) -\gamma
^{2}J_{qm}^{2}(\gamma r)\right] ,  \label{E2b1}
\end{eqnarray}%
are the parts induced by the presence of a single plate at $z=0$. Note that
the latter is finite for points away from the plate and the renormalization
is necessary for the boundary-free part only. The renormalized boundary-free
part is given by the following simple expression \cite{Beze07}:
\begin{equation}
\langle E^{2}\rangle _{\mathrm{ren}}^{(s)}=\langle B^{2}\rangle _{\mathrm{ren%
}}^{(s)}=-\frac{(q^{2}-1)(q^{2}+11)}{180\pi r^{4}}.  \label{E2s1}
\end{equation}%
The corresponding VEV is negative.

After the integration over $k$, the single plate-induced part is presented
in the form%
\begin{equation}
\langle E^{2}\rangle _{1}^{(b)}=-\langle B^{2}\rangle _{1}^{(b)}=\frac{2q}{%
\pi z^{4}}\sideset{}{'}{\sum}_{m=0}^{\infty }\int_{0}^{\infty }du\,u^{3}%
\left[ J_{qm}^{2}(ur/z)K_{0}(2u)+G_{qm}(ur/z)Q(2u)\right] ,  \label{E2b11}
\end{equation}%
with the notation%
\begin{equation}
Q(x)=K_{0}(x)+2K_{1}(x)/x,  \label{Qx}
\end{equation}%
where $K_{\nu }(x)$ is the modified Bessel function. As we see, the
boundary-induced part is positive for the electric field and negative for
the magnetic field. For a metallic plate in background of Minkowski
spacetime one has $q=1$ and the summation over $m$ in (\ref{E2b11}) can be
explicitly done. After the integration over $u$, we obtain the well known
result $\langle E^{2}\rangle _{1}^{(b)}|_{q=1}=3/(4\pi z^{4})$. For $r\ll z$
we use the asymptotic expression for the Bessel function for small values of
the argument \cite{hand}. In this way, to the leading order one finds: $%
\langle E^{2}\rangle _{1}^{(b)}\approx $ $q/(4\pi z^{4})$ for $q>1$. Note
that the next-to-leading order term is of the order $(r/z)^{2q-2}$. In this
regime the total VEV of the electric field squared is dominated by the
boundary-free part and it is negative. In the opposite limiting case, $r\gg
z $, to the leading order the boundary-induced VEV\ coincides with the
corresponding quantity in Minkowski bulk. In this limit the total VEV is
dominated by the boundary-induced part and it is positive for the electric
field. For the magnetic field the total VEV is negative everywhere.

\begin{figure}[tbph]
\begin{center}
\epsfig{figure=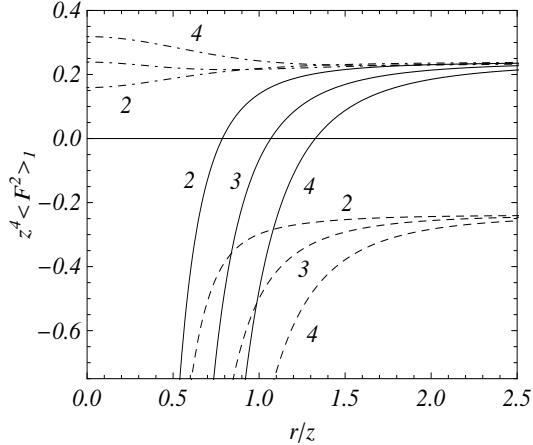,width=7.cm,height=6.cm}
\end{center}
\caption{VEV of the field squared, $z^{4}\langle F^{2}\rangle _{1}$, for the
electric (full curves) and magnetic (dashed curves) fields in the geometry
of a single metallic plate. The numbers near curves are the values of the
parameter $q$. The boundary-induced part in the VEV of the electric field
squared is plotted separately by the dot-dashed curves. }
\label{fig1}
\end{figure}

For the special case with integer values of the parameter $q$, the summation
over $m$ in (\ref{E2b11}) can be done by using the formulas \cite%
{Davi88,Beze07,Prud86}
\begin{eqnarray}
\sideset{}{'}{\sum}_{m=0}^{\infty }J_{qm}^{2}(y) &=&\frac{1}{2q}%
\sum_{l=0}^{q-1}J_{0}(2ys_{l}),  \notag \\
\sideset{}{'}{\sum}_{m=0}^{\infty }G_{qm}(y) &=&\frac{1}{2q}%
\sum_{l=0}^{q-1}\cos (2\pi l/q)J_{0}(2ys_{l}).  \label{Summq}
\end{eqnarray}%
with%
\begin{equation}
s_{l}=\sin (\pi l/q).  \label{sl}
\end{equation}%
In this case, for the single plate-induced parts we find:%
\begin{equation}
\langle E^{2}\rangle _{1}^{(b)}=-\langle B^{2}\rangle _{1}^{(b)}=\frac{1}{%
4\pi }\sum_{l=0}^{q-1}\frac{\left( 3-4s_{l}^{2}\right) z^{2}-r^{2}s_{l}^{2}}{%
\left( z^{2}+r^{2}s_{l}^{2}\right) ^{3}}.  \label{E2b1qint}
\end{equation}%
The $l=0$ term in this expression coincides with the corresponding VEV for a
plate in background of Minkowski spacetime: $\langle E^{2}\rangle
_{1}^{(b)}|_{q=1}=3/(4\pi z^{4})$. In figure \ref{fig1} we plot the VEVs for
the electric (full curves) and magnetic (dashed curves) field squared in the
geometry of a single conducting plate located at $z=0$ as a function of the
ratio $r/z$. The numbers near the curves are the corresponding values of the
parameter $q$. The dot-dashed curves correspond to the boundary-induced part
in the VEV of the field squared, $z^{4}\langle E^{2}\rangle _{1}^{(b)}$.
Note that, in general, the boundary-induced part $\langle E^{2}\rangle
_{1}^{(b)}$ is a non-monotonic function of $r$. The corresponding VEVs for a
plate in Minkowski spacetime would be presented by the horizontal lines $%
3/(4\pi )$ and $-3/(4\pi )$ for the electric and magnetic fields
respectively.

Now we turn to the second plate induced parts given by (\ref{F22}). By using
the expansion $(e^{y}-1)^{-1}=\sum_{n=1}^{\infty }e^{-ny}$, they can be
presented in the form:%
\begin{eqnarray}
\langle E^{2}\rangle _{2} &=&\frac{1}{a^{4}}C(r/a)+\frac{1}{a^{4}}%
\sum_{j=\pm 1}D(r/a,jz/a),  \notag \\
\langle B^{2}\rangle _{2} &=&\frac{1}{a^{4}}C(r/a)-\frac{1}{a^{4}}%
\sum_{j=\pm 1}D(r/a,jz/a),  \label{E2CD}
\end{eqnarray}%
with the functions%
\begin{eqnarray}
C(x) &=&-\frac{4q}{\pi }\sum_{n=1}^{\infty }\sideset{}{'}{\sum}%
_{m=0}^{\infty }\int_{0}^{\infty }d\gamma \gamma ^{3}\left[ G_{qm}(\gamma
x)Q(2n\gamma )-J_{qm}^{2}(\gamma x)K_{0}(2n\gamma )\right] ,  \notag \\
D(x,y) &=&\frac{2q}{\pi }\sum_{n=1}^{\infty }\sideset{}{'}{\sum}%
_{m=0}^{\infty }\int_{0}^{\infty }d\gamma \gamma ^{3}\left[ G_{qm}(\gamma
x)Q(2(n-y)\gamma )+J_{qm}^{2}(\gamma x)K_{0}(2(n-y)\gamma )\right] .
\label{CD}
\end{eqnarray}%
Note that the $z$-dependent parts for the VEVs of the electric and magnetic
fields have opposite signs. In particular, they are cancelled in the
corresponding energy density (see the next section). The quantity $\langle
E^{2}\rangle _{2}$ is finite on the plate at $z=0$ and diverges on the plate
at $z=a$. The divergence comes from the $n=1$ term in the expression for the
function $D(x,y)$. Note that the contribution of this term to the VEV of the
electric field squared coincides with (\ref{E2b11}) with the replacement $%
z\rightarrow a-z$. Hence, it presents the VEV induced by the plate at $z=a$
when the plate at $z=0$ is absent. Combining the results for the single
plate and second plate-induced parts, the total VEV is presented in the form%
\begin{eqnarray}
\langle F^{2}\rangle  &=&\langle F^{2}\rangle _{\mathrm{ren}}^{(s)}-\frac{4q%
}{\pi }\sum_{n=1}^{\infty }\sideset{}{'}{\sum}_{m=0}^{\infty
}\int_{0}^{\infty }d\gamma \gamma ^{3}\left[ G_{qm}(\gamma r)Q(2na\gamma
)-J_{qm}^{2}(\gamma r)K_{0}(2na\gamma )\right]   \notag \\
&+&\delta _{F}\frac{2q}{\pi }\sum_{n=-\infty }^{\infty }\sideset{}{'}{\sum}%
_{m=0}^{\infty }\int_{0}^{\infty }d\gamma \gamma ^{3}\left[ G_{qm}(\gamma
r)Q(2|na-z|\gamma )+J_{qm}^{2}(\gamma r)K_{0}(2|na-z|\gamma )\right] ,
\label{totF2}
\end{eqnarray}%
where $\delta _{E}=1$ and $\delta _{B}=-1$. The single plate parts in this
expression are presented by the $n=0$ and $n=1$ terms in the last summation
on the right-hand side. The VEV is not changed under the replacement $%
z\rightarrow a-z$, which is a direct consequence of the problem symmetry
with respect to the plane $z=a/2$.

\begin{figure}[tbph]
\begin{center}
\begin{tabular}{cc}
\epsfig{figure=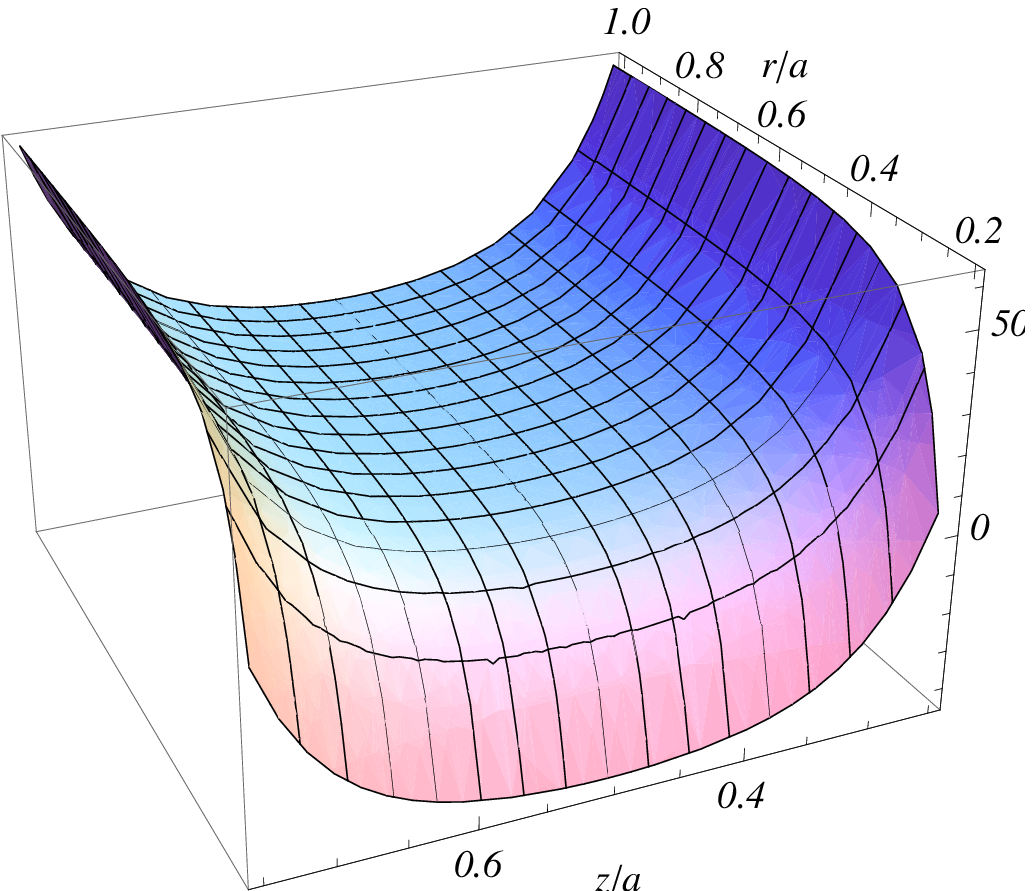,width=7.cm,height=6.cm} & \quad %
\epsfig{figure=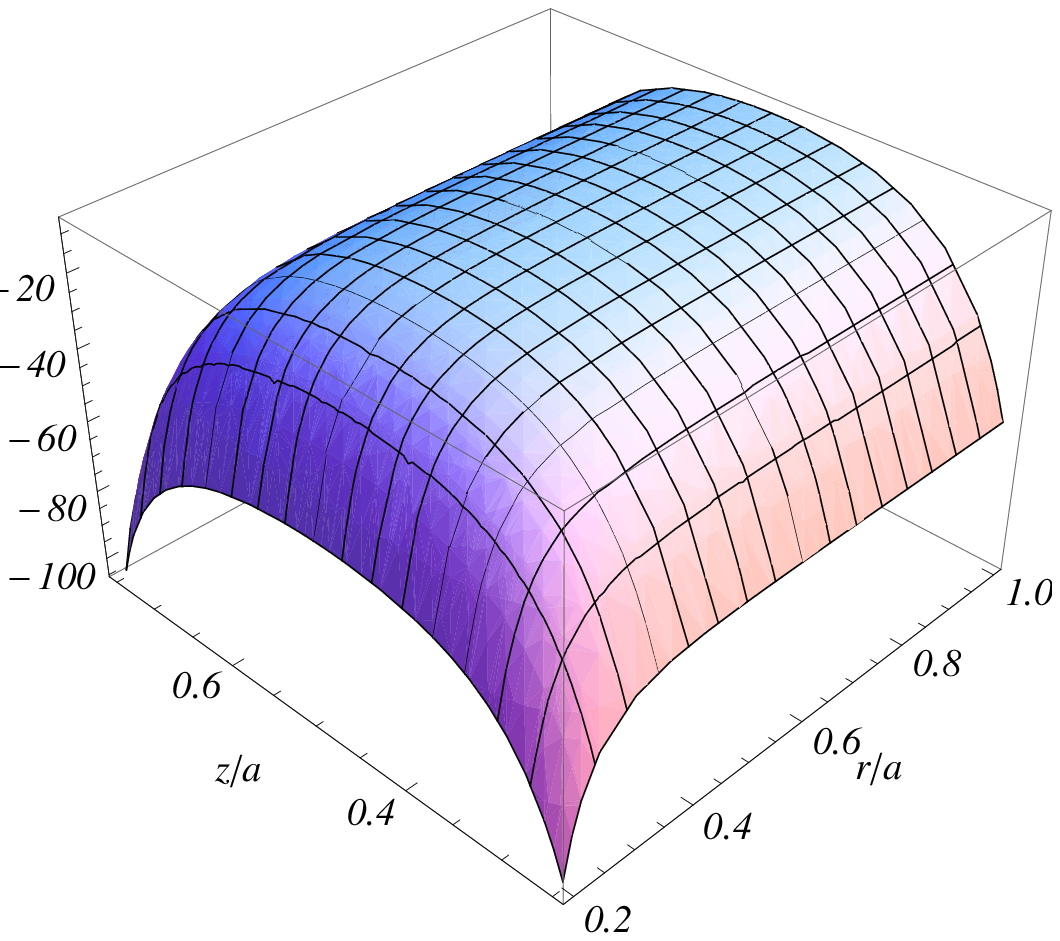,width=7.cm,height=6.cm}%
\end{tabular}%
\end{center}
\caption{VEVs of the electric field squared (left plot), $a^{4}\langle
E^{2}\rangle $, and the magnetic field squared (right plot), $a^{4}\langle
B^{2}\rangle $, in the region between two plates for $q=2$.}
\label{fig2}
\end{figure}

For integer values of the parameter $q$, by using formulas (\ref{Summq}),
one finds%
\begin{eqnarray}
C(x) &=&-\frac{1}{2\pi }\sum_{l=0}^{q-1}\sum_{n=1}^{\infty }\frac{\left(
1-4s_{l}^{2}\right) n^{2}+x^{2}s_{l}^{2}}{\left( n^{2}+x^{2}s_{l}^{2}\right)
^{3}},  \notag \\
D(x,y) &=&\frac{1}{4\pi }\sum_{l=0}^{q-1}\sum_{n=1}^{\infty }\frac{\left(
3-4s_{l}^{2}\right) (n-y)^{2}-x^{2}s_{l}^{2}}{\left[ (n-y)^{2}+x^{2}s_{l}^{2}%
\right] ^{3}}.  \label{CDintq}
\end{eqnarray}%
After the summation over $n$, the function $C(x)$ can also be presented in
the form%
\begin{equation}
C(x)=-\frac{1}{2\pi }\sum_{l=0}^{q-1}\left[ \left( 1-4s_{l}^{2}\right)
h_{2}(xs_{l})+4s_{l}^{2}h_{3}(xs_{l})\right] ,  \label{Cx}
\end{equation}%
with the notations%
\begin{eqnarray}
h_{2}(b) &=&\sum_{n=1}^{\infty }\frac{1}{\left( n^{2}+b^{2}\right) ^{2}}=-%
\frac{1}{2b^{4}}+\frac{\pi }{4b^{3}}\left[ \coth (\pi b)+\frac{\pi b}{\sinh
^{2}(\pi b)}\right] ,  \notag \\
h_{3}(b) &=&\sum_{n=1}^{\infty }\frac{b^{2}}{\left( n^{2}+b^{2}\right) ^{3}}%
=-\frac{1}{2b^{4}}+\frac{\pi }{16b^{3}}\left[ 3\coth (\pi b)+\frac{3\pi b}{%
\sinh ^{2}(\pi b)}+\frac{2\pi ^{2}b^{2}}{\sinh ^{2}(\pi b)}\coth (\pi b)%
\right] .  \label{h3}
\end{eqnarray}%
Note that $h_{2}(0)=\pi ^{4}/90$. For integer $q$, the expression of the
total VEV takes the form:%
\begin{eqnarray}
\langle F^{2}\rangle  &=&\langle F^{2}\rangle _{\mathrm{ren}}^{(s)}-\frac{1}{%
2\pi }\sum_{l=0}^{q-1}\sum_{n=1}^{\infty }\frac{\left( 1-4s_{l}^{2}\right)
n^{2}a^{2}+r^{2}s_{l}^{2}}{\left( n^{2}a^{2}+r^{2}s_{l}^{2}\right) ^{3}}
\notag \\
&&+\frac{\delta _{F}}{4\pi }\sum_{l=0}^{q-1}\sum_{n=-\infty }^{\infty }\frac{%
\left( 3-4s_{l}^{2}\right) (na-z)^{2}-r^{2}s_{l}^{2}}{\left[
(na-z)^{2}+r^{2}s_{l}^{2}\right] ^{3}}.  \label{totF2intq}
\end{eqnarray}%
In figure \ref{fig2} we display the dependence on $r/a$ and $z/a$ for the
VEVs of the electric (left plot) and magnetic (right plot) field squared, in
the region between the plates for a cosmic string with $q=2$. For the VEV
for the electric field squared the boundary-induced part dominates near the
plates and it is positive in this region. Near the string the VEV is
dominated by the boundary-free part and it is negative. The VEV of the
magnetic field squared is negative everywhere.

In the region $z<0$ the VEVs for the field squared are given by (\ref{F1dec}%
), where the plate induced part is still given by the expression (\ref{E2b1}%
). The same decomposition (\ref{F1dec}) is valid for the region $z>a$, with
the difference that now the single plate part is given by the expression (%
\ref{E2b1}) with the replacement $z\rightarrow z-a$.

The VEV of the electric field squared determines the
Casimir-Polder force acting on a polarizable particle. In the
static limit, when the dispersion of polarizability can be
neglected, the corresponding potential is given by the expression
\begin{equation}
U(r,z)=-\frac{1}{2}\alpha _{0}\langle E^{2}\rangle ,  \label{Ucp}
\end{equation}%
where $\alpha _{0}$ is the static polarizability. The Casimir-Polder force
in the boundary-free cosmic string geometry has been recently investigated
in \cite{Saha11} for the general case of anisotropic polarizability with
dispersion. In the isotropic case the corresponding force is repulsive with
respect to the cosmic string. From (\ref{E2b11}) it can be easily seen that $%
\partial _{z}\langle E^{2}\rangle _{1}^{(b)}<0$ and, hence, the $z$%
-component of the Casimir-Polder force in the geometry of a single plate is
attractive with respect to the plate. In the region between two plates, this
component vanishes at $z=a/2$ and it is attractive to the nearest plate for $%
z\neq a/2$. In the presence of the plates, the radial component of the total
Casimir-Polder force remains repulsive with respect to the string, thought
the boundary induced part separately can be either repulsive or attractive.

\section{Energy-momentum tensor and the Casimir force}

\label{sec:EMT}

In this section we consider the VEV of the energy-momentum tensor for the
geometry of two parallel metallic plates. This VEV is obtained from the
general formula (\ref{VEVbil}) considering the standard expression for the
energy-momentum tensor of the electromagnetic field. Substituting the mode
functions for the electric and magnetic fields, given in section \ref%
{sec:Inter}, in a way similar to that for the field squared, it can be seen
that the single plate parts in the VEV of the energy-momentum tensor vanish,
$\langle T_{\mu \nu }\rangle _{1}^{(b)}=0$. In the region between the plates
the total VEV is presented in the decomposed form%
\begin{equation}
\langle T_{\mu \nu }\rangle =\langle T_{\mu \nu }\rangle ^{(s)}+\langle
T_{\mu \nu }\rangle ^{(b)},  \label{EMTdec}
\end{equation}%
where the renormalized energy-momentum tensor for the pure string part is
given by the expression \cite{Frol87,Dowk87}
\begin{equation}
\langle T_{\nu }^{\mu }\rangle ^{(s)}=-\frac{(q^{2}-1)(q^{2}+11)}{720\pi
^{2}r^{4}}\text{diag}(1,1,-3,1).  \label{T00s}
\end{equation}%
The energy density in (\ref{T00s}) can be directly obtained from the
expressions for the electric and magnetic field squared. Providing the
energy density, the spatial components are obtained by using the zero trace
condition, $\langle T_{\mu }^{\mu }\rangle ^{(s)}=0$, and the covariant
continuity equation, $\nabla _{\mu }\langle T_{\nu }^{\mu }\rangle ^{(s)}=0$%
. For the boundary-induced parts in the separate components we get the
following expressions%
\begin{eqnarray}
\left\langle T_{0}^{0}\right\rangle ^{(b)} &=&-\frac{q}{\pi ^{2}}%
\sum_{n=1}^{\infty }\sideset{}{'}{\sum}_{m=0}^{\infty }\int_{0}^{\infty
}d\gamma \gamma ^{3}\left[ G_{qm}(\gamma r)Q(2an\gamma )-J_{qm}^{2}(\gamma
r)K_{0}(2an\gamma )\right] ,  \notag \\
\left\langle T_{1}^{1}\right\rangle ^{(b)} &=&-\frac{q}{\pi ^{2}}%
\sum_{n=1}^{\infty }\sideset{}{'}{\sum}_{m=0}^{\infty }\int_{0}^{\infty
}d\gamma \gamma ^{3}\left\{ J_{qm}^{\prime 2}(\gamma r)+\left[ 1-\left(
\frac{qm}{\gamma r}\right) ^{2}\right] J_{qm}^{2}(\gamma r)\right\}
K_{0}(2an\gamma ),  \notag \\
\left\langle T_{2}^{2}\right\rangle ^{(b)} &=&\frac{q}{\pi ^{2}}%
\sum_{n=1}^{\infty }\sideset{}{'}{\sum}_{m=0}^{\infty }\int_{0}^{\infty
}d\gamma \gamma ^{3}\left\{ J_{qm}^{\prime 2}(\gamma r)-\left[ 1+\left(
\frac{qm}{\gamma r}\right) ^{2}\right] J_{qm}^{2}(\gamma r)\right\}
K_{0}(2an\gamma ),  \notag \\
\left\langle T_{3}^{3}\right\rangle ^{(b)} &=&\frac{q}{\pi ^{2}}%
\sum_{n=1}^{\infty }\sideset{}{'}{\sum}_{m=0}^{\infty }\int_{0}^{\infty
}d\gamma \gamma ^{3}\left[ G_{qm}(\gamma r)Q(2an\gamma )+J_{qm}^{2}(\gamma
r)K_{0}(2an\gamma )\right] ,  \label{T33b}
\end{eqnarray}%
and the off-diagonal components vanish. It is easily checked that the
boundary-induced part is traceless. Moreover, it does not depend on the $z$%
-coordinate.

Note that, by using the relation
\begin{equation}
J_{qm}^{\prime 2}(\gamma r)+\left[ 1-\left( \frac{qm}{\gamma r}\right) ^{2}%
\right] J_{qm}^{2}(\gamma r)=\frac{2}{r^{2}}\int_{0}^{r}dx\,xJ_{qm}^{2}(%
\gamma x),  \label{relBes}
\end{equation}%
the expression for the radial stress may also be written in the form%
\begin{equation}
\left\langle T_{1}^{1}\right\rangle ^{(b)}=-\frac{2q}{\pi ^{2}r^{2}}%
\sum_{n=1}^{\infty }\sideset{}{'}{\sum}_{m=0}^{\infty }\int_{0}^{\infty
}d\gamma \gamma ^{3}K_{0}(2an\gamma )\int_{0}^{r}dx\,xJ_{qm}^{2}(\gamma x),
\label{T11b}
\end{equation}%
from which it follows that the boundary-induced part in the vacuum pressure
along the radial direction, $p_{1}^{(b)}=-\left\langle
T_{1}^{1}\right\rangle ^{(b)}$, is always positive. The same is the case for
the boundary-free part. Note that, from (\ref{T33b}) we also have $%
\left\langle T_{1}^{1}\right\rangle ^{(b)}+\left\langle
T_{2}^{2}\right\rangle ^{(b)}\leqslant 0$. It can be explicitly checked that
the boundary-induced parts obey the covariant continuity equation, $\nabla
_{\mu }\left\langle T_{\nu }^{\mu }\right\rangle ^{(b)}=0$, which for the
geometry under consideration is reduced to the single equation $\partial
_{r}(r\left\langle T_{1}^{1}\right\rangle ^{(b)})=\left\langle
T_{2}^{2}\right\rangle ^{(b)}$.

Let us consider the behavior of the boundary-induced part in the VEV\ of the
energy-momentum tensor near the string assuming that $r\ll a$. Assuming that
$q>1$ and by using the formula for the Bessel function for small arguments,
to the leading order we find%
\begin{equation}
\left\langle T_{0}^{0}\right\rangle ^{(b)}\approx \left\langle
T_{3}^{3}\right\rangle ^{(b)}\approx -\left\langle T_{1}^{1}\right\rangle
^{(b)}\approx -\left\langle T_{2}^{2}\right\rangle ^{(b)}\approx \frac{q\pi
^{2}}{720a^{4}}.  \label{EMTnear}
\end{equation}%
In particular, we see that near the string the boundary-induced part in the
energy density is positive. Note that in this region the total VEV is
dominated by the boundary-free part. At large distances from the string, $%
r\gg a$, the effects induced by the nontrivial topology of the string are
small and, to the leading order, the VEV coincides with the corresponding
expression for metallic plates in Minkowski spacetime \cite{Brow69}:%
\begin{equation}
\left\langle T_{\nu }^{\mu }\right\rangle \approx \left\langle T_{\nu }^{\mu
}\right\rangle ^{(b)}\approx \left\langle T_{\nu }^{\mu }\right\rangle _{%
\text{M}}=-\frac{\pi ^{2}}{720a^{4}}\text{diag}(1,1,1,-3).  \label{EMTfar}
\end{equation}%
In this region the both boundary-free and boundary-induced parts in the
vacuum energy density are negative.

\begin{figure}[tbph]
\begin{center}
\epsfig{figure=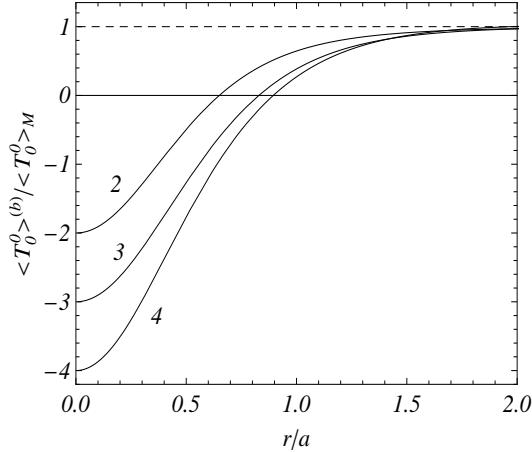,width=7.cm,height=6.cm}
\end{center}
\caption{The ratio of the boundary-induced part in the energy density to the
corresponding quantity for parallel plates in Minkowski spacetime. The
numbers near the curves are the values of the parameter $q$. }
\label{fig3}
\end{figure}

For integer values of $q$, by using the formulas (\ref{Summq}), one finds
the following expressions (no summation over $\mu $)%
\begin{equation}
\left\langle T_{\mu }^{\mu }\right\rangle ^{(b)}=-\frac{a^{-4}}{8\pi ^{2}}%
\sum_{l=0}^{q-1}\left[ f_{\mu ,2}(s_{l})h_{2}(rs_{l}/a)+f_{\mu
,3}(s_{l})h_{3}(rs_{l}/a)\right] ,  \label{T33qint2}
\end{equation}%
where%
\begin{eqnarray}
f_{0,2}(x) &=&1-4x^{2},\quad f_{0,3}(x)=4x^{2},\quad f_{1,2}(x)=1,\quad
f_{1,3}(x)=0,  \notag \\
f_{2,2}(x) &=&1,\quad f_{2,3}(x)=-4,\quad f_{3,2}(x)=4x^{2}-3,\quad
f_{3,3}(x)=4-4x^{2}.  \label{fmu2}
\end{eqnarray}%
The $l=0$ terms in (\ref{T33qint2}) coincide with the corresponding
quantities for parallel plates in Minkowski spacetime (see (\ref{EMTfar})).
In figure \ref{fig3} we present the ratio of the boundary-induced part of
the energy density to the corresponding quantity for parallel plates in
Minkowski spacetime as a function of the distance from the string. The
numbers near the curves are the values of the parameter $q$.

In the regions $z<0$ and $z>a$ the boundary-induced part in the VEV of the
energy-momentum tensor vanishes and in these regions $\langle T_{\mu \nu
}\rangle =\langle T_{\mu \nu }\rangle ^{(s)}$.

\begin{figure}[tbph]
\begin{center}
\epsfig{figure=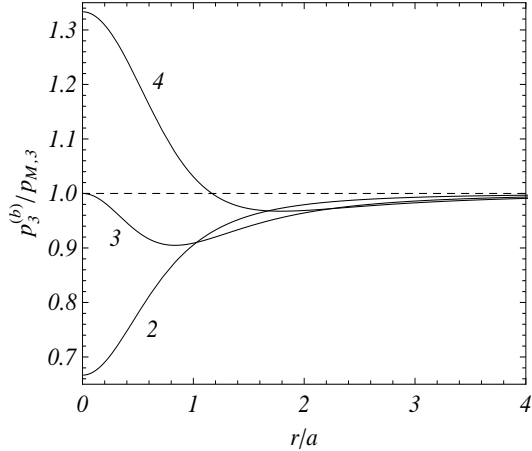,width=7.cm,height=6.cm}
\end{center}
\caption{The ratio of the Casimir stress on the plate to the corresponding
quantity in Minkowski spacetime as a function of the distance $r/a$ from the
string for separate values of the parameter $q$. }
\label{fig4}
\end{figure}

The Casimir force acting per unit surface of the plate is determined by the
component $\left\langle T_{3}^{3}\right\rangle $. For the corresponding
effective pressure one has: $p_{3}=-\left\langle T_{3}^{3}\right\rangle $.
The boundary free part of the pressure is the same for both sides of the
plate and it does not contribute to the net force. Hence, the force per unit
surface of the plate is determined by the boundary induced part of the
pressure along the $z$-direction:%
\begin{equation}
p_{3}^{(b)}=-\frac{q}{\pi ^{2}a^{4}}\sum_{n=1}^{\infty }\sideset{}{'}{\sum}%
_{m=0}^{\infty }\int_{0}^{\infty }d\gamma \gamma ^{3}\left[ G_{qm}(\gamma
r/a)Q(2n\gamma )+J_{qm}^{2}(\gamma r/a)K_{0}(2n\gamma )\right] .
\label{Casp}
\end{equation}%
Unlike to the case of Minkowski bulk, the Casimir stress on the plates is
not uniform. The effective pressure (\ref{Casp}) is always negative and,
hence, the corresponding Casimir force is always attractive. For integer
values of the parameter $q$ we have:%
\begin{equation}
p_{3}^{(b)}=-\frac{a^{-4}}{8\pi ^{2}}\sum_{l=0}^{q-1}\left[ \left(
3-4s_{l}^{2}\right) h_{2}(rs_{l}/a)-4\left( 1-s_{l}^{2}\right)
h_{3}(rs_{l}/a)\right] ,  \label{Caspqint}
\end{equation}%
with $s_{l}$ defined by (\ref{sl}). The $l=0$ term in this expression
coincides with the corresponding quantity for plates in Minkowski spacetime:
$p_{\text{M},3}=-\pi ^{2}a^{-4}/240$. In figure \ref{fig4} we plot the ratio
$p_{3}^{(b)}/p_{\text{M},3}$ as a function of the distance from the string
(in units of the separation between the plates) for separate values of the
parameter $q$ (numbers near the curves).

\section{Conclusion}

\label{sec:Conc}

In this paper we have investigated the combined effects from nontrivial
topology and boundaries on properties of the electromagnetic vacuum. The
nontrivial topology of the spacetime is induced by the presence of the
cosmic string, and the for the boundaries we have considered the classical
Casimir geometry with two parallel conducting plates. Among the most
important characteristics of the vacuum state are the VEVs of the electric
and magnetic fields squared and the energy-momentum tensor. In order to
evaluate these VEVs we have employed the direct mode summation technique. By
applying the Abel-Plana summation formula to the mode sum over the
eigenvalues of the wave vector component along the cosmic string, we have
explicitly decomposed the VEVs into the pure string and boundary-induced
parts. The pure string parts in the VEVs of the electric and magnetic fields
squared coincide and they are negative everywhere. The presence of the
cosmic string increases the boundary-induced parts in the VEVs of the field
squared when compared with the Minkowski spacetime results.

The boundary part in the VEVs of the field squared are further split into
the single plate and the second plate-induced parts. Single plate parts are
given by expression (\ref{E2b11}) and they have opposite signs for the
electric and magnetic fields. For points near the string, $r\ll z$, the VEV
of the electric field squared is dominated by the pure string part and it is
negative. Near the plate, $z\ll r$, the plate induced part dominates and the
VEV is positive. The VEV of the magnetic field squared is negative
everywhere. The second plate-induced parts are presented in the form (\ref%
{E2CD}) with the functions defined in (\ref{CD}). The $z$-dependent parts in
the expressions for the electric and magnetic fields squared have opposite
signs and they are cancelled in the expression for the vacuum energy
density. The general formulas are simplified in a special case of integer
values of the parameter $q$. The corresponding expressions take the form (%
\ref{E2b1qint}) and (\ref{CDintq}). The VEV of the electric field squared
determines the Casimir-Polder force on a polarizable particle. In the
geometry of a single plate the force is attractive with respect to the plate
and repulsive with respect to the string. In the region between two plates,
the Casimir-Polder force is attractive to the nearest plate and remains
repulsive with respect to the string.

The boundary-induced part in the VEV of the energy-momentum tensor for the
electromagnetic field is different from zero in the region between the
plates only. The expressions for the separate components are given by
formulas (\ref{T33b}). The VEVs are uniform with respect to the coordinate
along the cosmic string and depend on the distance from the string only. The
boundary-induced part in the vacuum energy density is positive near the
string and negative at large distances from the string. The radial effective
pressure is always positive. The general formulas for the VEV of the
energy-momentum tensor are simplified for integer values of the parameter $q$%
. The corresponding expressions are given by (\ref{T33qint2}). Unlike to the
geometry of two conducting plates in Minkowski spacetime, in the geometry
with cosmic string the Casimir stresses on the plates are non-uniform. They
depend on the distance from the string and this dependence is not monotonic.
The Casimir forces acting on the plates are always attractive. They are
given by (\ref{Casp}) for the general case and by (\ref{Caspqint}) for
integer values of the parameter $q$.

\section*{Acknowledgments}

ERBM thanks Conselho Nacional de Desenvolvimento Cient\'{\i}fico e Tecnol%
\'{o}gico (CNPq) for partial financial support. AAS was supported by
PVE/CAPES Program.

\end{document}